\documentclass[15pt a4paper]{article}

\usepackage{latexsym}
\usepackage{amsmath,amssymb,amsfonts,amsthm}
\usepackage{harvard_local}
\usepackage{graphicx}

\graphicspath{{./}{./figs/}}

\begin{document}

\title{Synchronizability and synchronization dynamics of weighed \\ and unweighed scale free networks with degree mixing}
\author{Mario di Bernardo, Franco Garofalo, Francesco Sorrentino\thanks{Corresponding author, Email:
fsorrent@unina.it} \thanks{The authors are listed in alphabetical
order}}

\date{Department of Systems and Computer Science, \\University of Naples Federico II,\\ Napoli, 80125, Italy.\\
} \maketitle

\begin{abstract}
We study the synchronizability and the synchronization dynamics of
networks of nonlinear oscillators. We investigate how the
synchronization of the network is influenced by some of its
topological features such as variations of the power law exponent
$\gamma$ and the degree correlation coefficient $r$. Using an
appropriate construction algorithm based on clustering the network
vertices in $p$ classes according to their degrees, we construct
networks with an assigned power law distribution but changing degree
correlation properties. We find that the network synchronizability
improves when the network becomes disassortative, i.e. when nodes
with low degree are more likely to be connected to nodes with higher
degree.  We consider the case of both weighed and unweighed
networks. The analytical results reported in the paper are then
confirmed by a set of numerical observations obtained on weighed and
unweighed networks of nonlinear R\"ossler oscillators. Using a
nonlinear optimization strategy we also show that negative degree
correlation is an emerging property of networks when
synchronizability is to be optimized. This suggests that negative
degree correlation observed experimentally in a number of physical
and biological networks might be motivated by their need to
synchronize better.
\medskip

\end{abstract}

\section{Introduction}
\bibliographystyle{agsm_local}
Networks of oscillators abound in physics, biology and engineering.
Examples include communication networks, sensor networks, neuronal
connectivity networks, biological networks and food webs
\cite{report}. Under certain conditions such networks are known to
synchronize on a common evolution, with all the oscillators
exhibiting the same asymptotic trajectory. Synchronization was
observed to play an important role in a wide variety of different
problems  (physical, ecological and physiological networks to name
just a few); see for example
\cite{KuraBOOK,WinBOOK,Er:Ko1,Er:Ko2,Wang,exploring,Str1,Str2}.

In this paper, we consider a network consisting of $N$ identical
oscillators coupled through the edges of the network itself
\cite{Ba:Pe02,Ni:Mo}. We suppose each oscillator is characterized
by its own dynamics, $x(t)=\{ x_i(t), i=1...N\}$, described by a
nonlinear set of ODEs, $\dot x = f(x)$. The dynamics of each
oscillator in the network is perturbed by the output function of
its neighbors represented by another nonlinear term, say $h=h(x)$.
The equations of motion for each oscillator can then be given as
follows:
\begin{equation}
\frac{dx_i}{dt}=f(x_i)-\sigma \sum_{j=1}^{N} {\mathcal L}_{ij}
h(x_j), \qquad i=1,2,...N, \label{eq:net}
\end{equation}
where $\sigma$ represents the overall strength of the coupling. Note
that information about the network topology is entirely contained in
the matrix $\mathcal L$, whose entries $\mathcal{L}_{ij}$, $j \neq
i$, are negative (zero) if node $i$ is (not) connected to node $j$,
with $\mid\mathcal{L}_{ij} \mid$ giving a measure of the strength of
the interaction.

In recent years, the analysis of large sets of data led to the
identification of some important structural properties  of many
real-world networks; see for a review \cite{report}. Among these,
the degree distribution $P(k)$, with the \emph{degree} $k$ being the
number of connections at a given node, was shown to be one of the
most important features. In particular, scale-free networks, which
are characterized by a power law degree distribution $P(k) \sim
k^{-\gamma}$, have been observed to be widely spread in nature
\cite{Am:Sc00}, \cite{Ba:Al99}. The main feature of scale free
networks is an high heterogeneity in the degree distribution (higher
than in purely random networks). Heuristically, this corresponds to
the fact that real networks have often many low-degree nodes and
only few nodes, termed as \emph{hubs}, with many connections (thus
leading to the high heterogeneity in the degree distribution). Such
a characterization is particularly relevant since it has been shown
to have important effects on the dynamics occurring over the
network, such as the spreading of epidemics \cite{New02epid},
\cite{Pa:Ve00a} or the distribution of packets in traffic dynamics
over the network \cite{Korea}, \cite{Be03}, \cite{Ar:dB:So}. Also,
it was shown that the scale-free structure of a network can have an
important effect on its synchronizability (see Sec.
\ref{sec:topology} for further details).

Recently, it has been suggested that real networks can often exhibit
other important features. For instance, in the real world, nodes are
not only abstract objects without any attribution, but each of them
is characterized by some intrinsic features. Examples of interest
could be the vertices age, spatial location, functional importance,
level of activity and the number of connections each vertex has
(that is the degree). An important topological property of physical
and biological networks is that often their nodes show preferential
attachment to other nodes in the network according to their degree
\cite{New02Ass}, \cite{New03Mix}. Networks are said to exhibit {\em
assortative} mixing (or positive correlation) if nodes of a given
degree tend to be attached with higher likelihood to nodes with
similar degree. (Similarly {\em disassortative} networks are those
with nodes of higher degree more likely to be connected to nodes of
lower degree.)

The presence of correlation has been detected experimentally in many
real-world networks. Interestingly, from the analysis of real
networks, it was noticed that social networks are characterized by
positive degree correlation, while physical and biological networks
show typically a disassortative structure \cite{New02Ass}. For
example, in \cite{Va:PaCORR}, Internet was found to exhibit
disassortative mixing at the Autonomous System level.

In mathematical terms, degree correlation can be quantified by means
of the observable $r$, introduced in \cite{New02Ass} and
\cite{New03Mix}. In particular, the coefficient $r$ was proposed in
\cite{New03Mix} as a normalized measure of degree correlation,
defined as the Pearson statistic:
\begin{equation}
\label{eq:r} r =\frac{\sum_{i j} ij (e_{ij} - q_i
q_j)}{{\sigma_q}^2},
\end{equation}
where $q_k$ is the probability that a randomly chosen edge is
connected to a node having degree $k$, $\sigma_q$ is the standard
deviation of the distribution $q_k$ and $e_{ij}$ represents the
probability that two vertices at the endpoints of a generic edge
have degree $i$ and $j$ respectively.

The main aim of this paper is to investigate the relationship
between the presence of degree correlation and the network
synchronizability properties. Specifically, we shall seek to
characterise how the presence of correlation on the network
affects the synchronizability of the nonlinear oscillators at the
nodes. We will show that correlation has indeed an effect on
synchronizability in both weighed and unweighed networks. Our main
finding is that disassortative networks  synchronize better. Thus,
as will be further discussed in the paper, the presence of
negative degree correlation often detected in technological and
particularly biological networks of nonlinear oscillators might be
motivated by its benefits in terms of the network
synchronizability.

The rest of the paper is organised as follows. In Sec. 2 we briefly
review the Master Stability Function approach and the effects on
synchronizability of heterogeneity in the degree distribution, which
is typical of scale free networks. In Sec.~3, we analyze how degree
correlation can influence the synchronizability of unweighed
networks and give an analytical explanation of the observed
phenomena. Also, an optimization approach is used to motivate the
conjecture that negative degree correlation can be an emerging
property of networks when the aim is to optimize their
synchronization. The case of weighed networks is discussed in Sec.
4, while the study of synchronization of both weighed and unweighed
networks of R\"ossler oscillators is presented in Sec. 5.

\section{Effects of Topology on the Network Synchronizability}
\label{sec:topology}
\subsection{The Master Stability Function Approach}
Recently, it has been proposed that the network topology, i.e. the
way in which the oscillators are mutually coupled among themselves,
has an important effect on the network \emph{synchronizability}.
This is defined in terms of  the linear stability of the synchronous
equilibrium ($x_i=x_2=...=x_N=x_s $), whose existence is guaranteed
by the fact that ${\mathcal L}$ is a zero row-sum matrix.

The stability of the synchronization manifold can be investigated by
perturbing trajectories lying on it along directions which are
orthogonal to the manifold itself. Namely, we consider sufficiently
small perturbations $\epsilon_i$ from the synchronous state so that
$x_i(t)=x_s(t)+\epsilon_i(t)$, and:

\begin{equation}
\frac{d \epsilon_i}{dt}= Jf(x_s) \epsilon_i-\sigma \sum_{j=1}^{N}
{\mathcal L}_{ij} Jh(x_s) \epsilon_j, \label{error}
\end{equation}
where $J$ denotes the Jacobian matrix. Following the strategy
presented in \cite{Pe:Ca}, Eq. (\ref{error}) can be transformed in
blocks of the form:
\begin{equation}
\frac{d \eta}{dt}= [Jf(x_s)-\sigma \lambda_i Jh(x_s)] \eta, \quad
i=1,2,...,N \label{blocks}
\end{equation}
where $\eta$ is defined in \cite{Pe:Ca} in terms of the
perturbations $\epsilon_i$, $\lambda_i$ are the real eigenvalues of
the coupling matrix $\mathcal{L}$, ordered in such a way that
$\lambda_1 \leq \lambda_2 ... \leq \lambda_N$ (in this manuscript we
will deal only with real eigenvalues of the matrix $\mathcal{L}$,
although under general conditions, there may be complex conjugate
eigenvalues). Note that $\lambda_1$ is structurally equal to 0, as
the corresponding eigenvector is associated to the mode lying within
the synchronization manifold. It is worth observing that using
(\ref{blocks}), the problem of studying the stability of a generic
complex network in (1) has been replaced by that of considering the
stability of $N$ simpler independent systems.

This approach is used in \cite{Pe:Ca} to derive the so-called {\em
Master Stability Function} (MSF). Specifically, eq. (\ref{blocks})
is considered as a parametric equation in a parameter $\alpha$,
given by:
\begin{equation}
\frac{d \eta}{dt}= [Jf(x_s)-\alpha Jh(x_s)] \eta \label{parametric}
\end{equation}
and the values of $\alpha$ are sought corresponding to the maximum
Lyapunov exponent of the system in (\ref{parametric}) being
negative.

Interestingly, it can be shown that for a broad class of  systems
(associated to different dynamic functions $f$ and output
functions $h$ ),
 the Master Stability Function is negative in a bounded range of the
parameter $\alpha$, say [$\alpha_{min},\alpha_{max}$].

Thus, in order to guarantee the stability of the synchronization
manifold, all the $\sigma \lambda_i$, for $i=2,3,...,N$, must lie in
the range [$\alpha_{min},\alpha_{max}$]. Simply, this condition
reduces to the following:
\begin{eqnarray}
\sigma \lambda_2 > \alpha_{min} \qquad \sigma \lambda_N <
\alpha_{max}, \label{eq:5}
\end{eqnarray}
or equivalently, to:
\begin{equation}
\frac{\lambda_N}{\lambda_2} < \frac{\alpha_{max}}{\alpha_{min}},
\label{eq:6}
\end{equation}
which guarantees the existence of at least one value of $\sigma$,
for which the synchronization manifold is stable. Note that while
$\alpha_{max}$ and $\alpha_{min}$ are completely determined by
assigning  the functions $f$ and $h$, $\lambda_N$ and $\lambda_2$
depend solely on the network topology. Thus, the synchronizability
of a given network  can be defined independently from the functional
form of the dynamical systems at its nodes (i.e. of the functions
$f$ and $h$).

In general, (\ref{eq:5}) defines a bounded range of values of
$\sigma$, say $I_{\sigma}=[\frac{\alpha_{min}}{\lambda_2},
\frac{\alpha_{max}}{\lambda_N}]$, for which  synchronization is
attained. Therefore, an  increase of $\lambda_2$ and a decrease of
$\lambda_N$ can lead to a larger interval $I_{\sigma}$. Thus,
minimizing the eigenratio $R=\lambda_N/\lambda_2$ yields a
broadening of the range of values of $\sigma$ over which the network
synchronizes.

Using the MSF approach, it is possible to investigate the effects on
synchronization of the structural properties of the network topology
(in terms of the eigenratio $\lambda_N/\lambda_2$). Hence, it is
meaningful to characterise how the network topological features
affect the Laplacian eigenratio.

\subsection{Synchronizability of Scale-free networks}
Sofar, the effects of the network topology on its synchronizability
have been studied mainly with respect to the presence of scale free
topologies or patterns in the network; see for example,
\cite{Ni:Mo}, \cite{paradox}, \cite{Fan:Wang:04},
\cite{Lu:Chen:Cheng}, \cite{Wu:03}, \cite{Mo:synchro},
\cite{Mo:synchro2}, \cite{Bocc1,Bocc2}.

Scale free networks, which are common in nature, were found to show
better synchronizability for increasing values of the power law
exponent in \cite{Ni:Mo}, \cite{paradox}. Specifically, the
relationship was analyzed between the network structure and its
synchronizability. An interesting phenomenon was observed which was
termed as the "paradox of heterogeneity"; specifically, although
heterogeneity in the degree distribution leads to a reduction in the
average distance between nodes (the so
called \emph{small world} effect \cite{Wa:St98}), 
it may suppress synchronization.

To explain the observed phenomena, in \cite{Ni:Mo}, the transition
of the underlying network from scale free (power law distributed) to
random (Poisson distributed) was shown to have a big impact on the
eigenratio $R$ of the Laplacian eigenvalues. Namely, a decrease of
the heterogeneous nature of the network was discovered to yield, as
a result, a reduction of $R$, thus increasing the synchronizability
of the network itself.

\section{Effects of degree correlation on the Laplacian eigenratio}

In \cite{IJBC_NDES}, we proposed that a further decrease of $R$ can
be detected when negative degree correlation is introduced among the
network nodes. Specifically, in \cite{IJBC_NDES}, using the
\emph{configuration model} \cite{Mo:Re95}, disassortative networks
were found to synchronize better. In this section,  we start by
extending some of the results presented therein to the case of
unweighed networks constructed by using the \emph{static model},
which was firstly introduced in \cite{Korea}.

\subsection{Network Construction methodology}

In this paper, in order to construct networks characterized by a
given degree distribution, we will use the following algorithm:

\begin{enumerate}
  \item We
associate to each vertex $i=1...\mathcal{N}$ a weight or fitness
$\phi_i \propto (i+\theta)^{-\alpha}$, with $\alpha=[0,1)$, $\theta$
being a fixed parameter.
  \item Then, links are added among the network
nodes with probability proportional to the vertices weights, until
$\mathcal{E}$ links have been introduced.
\end{enumerate}

Note that by using this methodology, the expected degree at each
node $i$, is $k_i \simeq \phi_i \mathcal{E}$, and it can be shown
that the degree distribution is expected to follow a power law,
$P(k) \sim k^{-\gamma}$, with $\gamma={{1+\alpha}\over{\alpha}}$.

Observe that the nodes \emph{fitness} $\phi_i$, is the only
parameter of the model; we wish to emphasize that assigning the
fitness corresponds to assigning the expected degree at the
vertices, thus the degree at the nodes can be considered  as an
equivalent parameter of the model.

\begin{figure}[t]
\centering
\begin{picture}(0,0)(0,0)
\put(125,-5){ ${\gamma}$} \put(-10,100){\large
$\frac{N}{\mathcal{N}}$}
\end{picture}
\includegraphics[width=8cm]{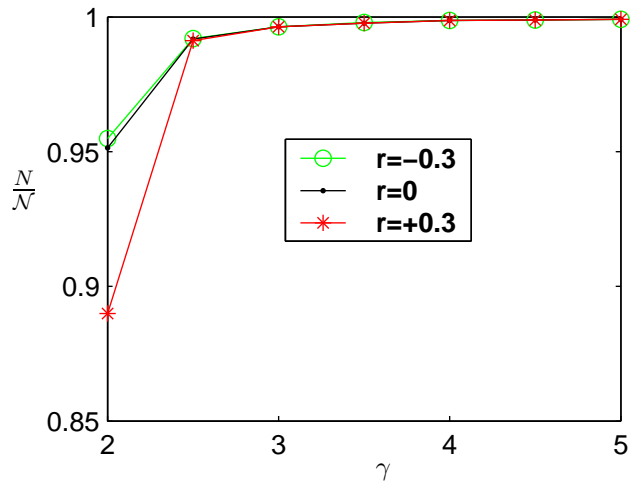}
 \caption{$N/\mathcal{N}$ as varying the degree distribution exponent
 $\gamma$ in networks characterized by different values of the
 degree correlation coefficient $r$. The legend is as follows: $r=-0.3$ (o), $r=0 (\cdot)$,
 $r=+0.3(\triangle)$. $\mathcal{N}=10^3$
nodes, $\mathcal{E}=4 \cdot 10^3$, $\theta=10$. \label{JC}}
\end{figure}

By using such a model, the giant component of the network may not
include all the network nodes, i.e the number of connected nodes,
say $N$, may be  $\leq \mathcal{N}$. In Fig. \ref{JC}, $N/{\mathcal
N}$ has been plotted as
varying both $\gamma$ and $r$. 
In what follows, we will consider only the largest connected part
of each generated network and refer to $N$ as the number of nodes
belonging to the giant component.

Using this methodology to construct a network, we can handle the
transition from very heterogenous scale free networks (when
$\alpha=1$ and $\gamma=2$), to highly homogenous ones (as $\alpha$
approaches 0 and $\gamma$ approaches $\infty$); note that in the
limit of $\gamma \rightarrow \infty$, in which to each vertex is
associated the same weight, we recover the classical \emph{random
network model}, which was introduced in \cite{Er:Re}. Moreover, this
does not imply a dependence of the total number of edges on the
degree distribution exponent $\gamma$ (as is the case of the widely
used \emph{ configuration model} \cite{Mo:Re95}).

By using this model we are able to reproduce networks
characterized by: (i) a given number of nodes $N$, (ii) a desired
number of edges $\mathcal{E}$, (iii) a degree distribution with a
desired value of  the exponent $\gamma$ 
(note that the same model has been already employed to study the
synchronization properties of networks in \cite{Lee05} and in
\cite{Oh:Rho}.) Then it is possible to devise a strategy similar to
the one presented in \cite{New03Mix}, to generate networks with a
given level of assortativity, i.e. a desired value of the degree
correlation coefficient $r$.

Notice that the model introduced in \cite{Korea} is known to exhibit
spontaneous negative degree correlation \cite{fermionic} when $2 <
\gamma <3$ (this phenomenon being referred to in \cite{fermionic} as
a \emph{fermionic constraint}, due to the impossibility of
connecting pairs of nodes with more than one link). The same
phenomenon has been observed also in networks constructed by using
the \emph{configuration model}, as described in
\cite{Par:New03,Ca:Ho01,uncorrelated}. However, in what follows, we
will disregard this particular effect; namely, we will compare
networks characterized by different degree correlation properties
according to the values of the index $r$ (including the case $r=0$),
which have indeed been \emph{measured} over the networks.

~~~~~~~~~~~~~~~~~~~~~~~~~~~~~~~~~~~~~~~~~~~~~~~~~~~~~~~~~~~~~~~~~~~~~~~~~~~~~~~~
~~~~~~~~~~~~~~~~~~~~~~~~~~~~~~~~~~~~~~~~~~~~~~~~~~~~~~~~~~~~~~~~~~~~~~~~~~~~~~~~

\subsection{Unweighed networks}

In the rest of this section we will investigate the case of
unweighed network topologies, with all the links being associated to
a constant unitary weight. Specifically in such a case, the matrix
$\mathcal{L}$ can be written as follows:
\begin{equation}
\label{ucoupling} \mathcal{L}=D-A,
\end{equation}
where $D=\{D_{ij}\}$ is a diagonal matrix such that $D_{ii}=k_i$,
$i=1,2,...,N$, and $A$ is the associated adjacency matrix. The
case of weighed topologies will be considered in Sec. 4.

\begin{figure}[b!]
\centering
\includegraphics[width=13cm]{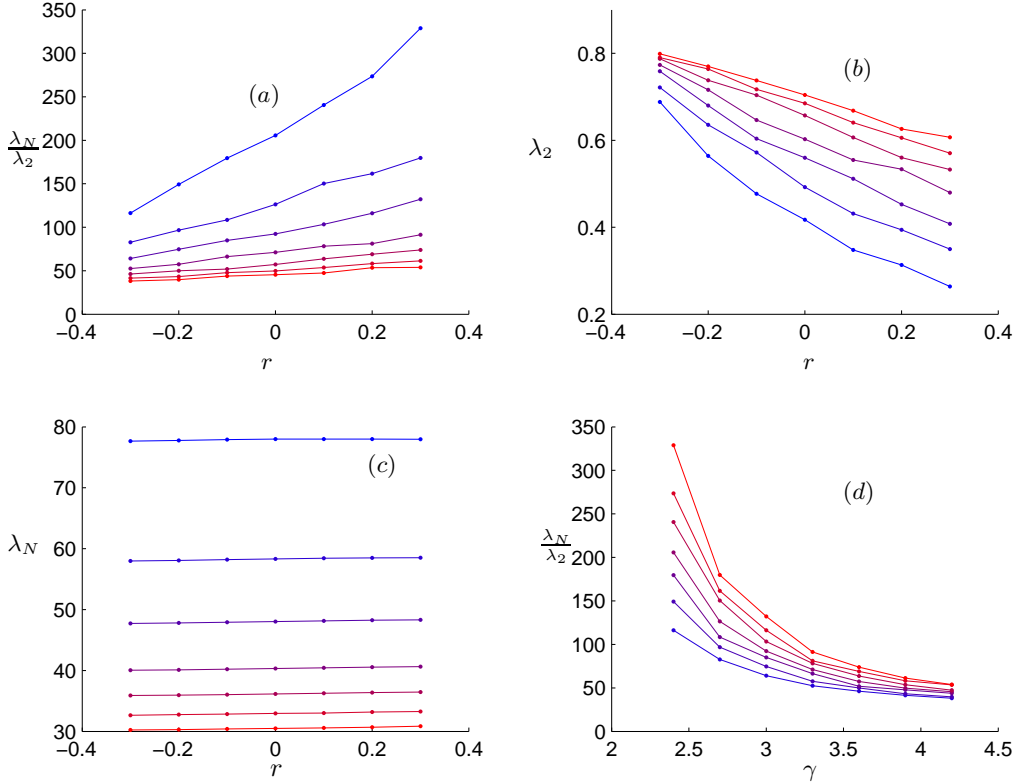}
\begin{picture}(0,0)(0,0)
\put(-290,-5){ ${r}$} \put(-290,149){\small $r$}
\put(-386,80){\small $\lambda_N$} \put(-390,230){
$\frac{\lambda_N}{\lambda_2}$} \put(-85,-5){\small ${\gamma}$}
\put(-85,149){\small $r$} \put(-185,80){\small
$\frac{\lambda_N}{\lambda_2}$} \put(-189,230){\small $\lambda_2$}
\put(-295,250){\small ${(a)}$} \put(-250,110){\small ${(c)}$}
\put(-70,260){\small ${(b)}$} \put(-70,100){\small ${(d)}$}
\end{picture}
\caption{\small Synchronizability of
degree correlated scale free networks of size $10^3$nodes. 
Behavior of the eigenratio $\lambda_{N}/\lambda_{2}$ (a), of the
second lowest eigenvalue $\lambda_2$ (b) and of the highest one
$\lambda_N$ (c), as functions of the correlation coefficient $r$
defined in [11], for $\gamma$ varying from 2.4 (top line) to 4.2
(bottom line) in steps of 0.3. (d) The eigenratio as function of
$\gamma$ as varying the correlation coefficient $r$  from $-0.3$
(bottom line) to $0.3$ (top line) in steps of $0.1$. 
The lines are guided by the eye. \label{fig:R}}
\end{figure}

The overall results are summarized in Fig. \ref{fig:R}(a) where
the effects of varying the degree correlation 
on the Laplacian eigenratio $R$ are shown for different values of
the degree distribution exponent $\gamma$. As discussed in
\cite{Ni:Mo}, in the case of uncorrelated networks ($r=0$),
synchronizability improves for increasing values of $\gamma$.
Moreover, for all values of $\gamma$, we observe a reduction of $R$
for decreasing values of $r$. This means that disassortative mixing
enhances the network synchronizability. Interestingly, as depicted
in Fig. \ref{fig:R}(b) and Fig. \ref{fig:R}(c),  we observe that,
under variations of the correlation parameter, the changes in $R$
seem to be mainly due to variations of $\lambda_2$. 
This is not surprising if we consider that $\lambda_N$ is known to
scale with $k_{max}$ \cite{Mohar}, the maximum degree at the
vertices 
and $k_{max}$
cannot vary with the degree correlation coefficient $r$.

It is worth noting here that the results provided in this section
are obtained by using a different network generation model from that
used in \cite{IJBC_NDES}. Moreover, the results shown in this paper
(and in particular those in Fig. 2) are found to confirm those
previously presented therein.

\subsection{An analytical explanation}

Here we briefly expound in this context the analysis provided in
\cite{IJBC_NDES}, in order to provide an explanation of the
behavior of the eigenratio $R$ as a function of the degree
correlation coefficient $r$ depicted in Fig. 2. 

Suppose the network vertices are divided in $p$ classes according to
their degree, i.e. clustering in each class ${\mathcal{C}}_i$ all
the vertices having degree $k_i$, ($i=1,2,...p$). Let us term as
$n_i$ the number of vertices in class ${\mathcal{C}}_i$, then the
probability of finding a vertex belonging to class ${\mathcal{C}}_i$
at the end of a randomly chosen edge within
the network is given by $q_i=n_i k_i / \sum_i n_i k_i$. 
In so doing, following \cite{New02Ass}, the presence of degree
correlation
 can be estimated by using the coefficient $r$ defined as:
\begin{equation}
\label{tre} r =\frac{{\mathbf{k}}^T ({\mathbf{E}} - {\mathbf{q}}
{\mathbf{q}}^T) \mathbf{k}}{ {\sigma^2_q}},
\end{equation}
where $\sigma_q$ is the standard deviation of the distribution
$q_i$, ${\mathbf{k}}={({k_1,  k_2, ..., k_p})^{T}}$,
${\mathbf{q}}={({q_1,  q_2, ..., q_p})^{T}}$ and
${\mathbf{E}}=\{e_{ij}\} \in {\mathbb{R}}^{{p} \times {p}}$, with
$e_{ij}$ being the probability that a randomly chosen edge in the
network connects nodes having degree $i$ and $j$.

From (\ref{tre}), it is possible to obtain the distribution of
edges among the network vertices as a function of $r$ as follows:
\begin{equation}
\label{EE} {\mathbf{E}}={\mathbf{q}\mathbf{q^T}}+ {r} {\sigma^2_q}
\mathbf{M},
\end{equation}
where $\mathbf{M}$ is a symmetric matrix having all row sums equal
to zero and appropriately normalized such that $\mathbf{k^T}
\mathbf{M} \mathbf{k} =1$. Specifically, we can express
${\mathbf{M}}$ as follows:
\begin{eqnarray*}
{\mathbf{M}}=\frac{{\mathbf{m}\mathbf{m}}^T}{({\mathbf{k}}^T {\mathbf{m}})^2},  
\end{eqnarray*}
where ${\mathbf{m}}={({m_1,  m_2, ..., m_p})^{T}}$ is a vector such
that $\sum_i m_i=0$ (for instance we can choose $m$ such that
$m_i \leq m_{i+1}$ for $i=1,2...,p-1$ 
in order to have a convenient form of the matrix $\mathbf{M}$ with
positive values near the main diagonal and negative values far away
from it). 
%
\begin{figure}[t]
\begin{center}
\includegraphics[width=7cm]{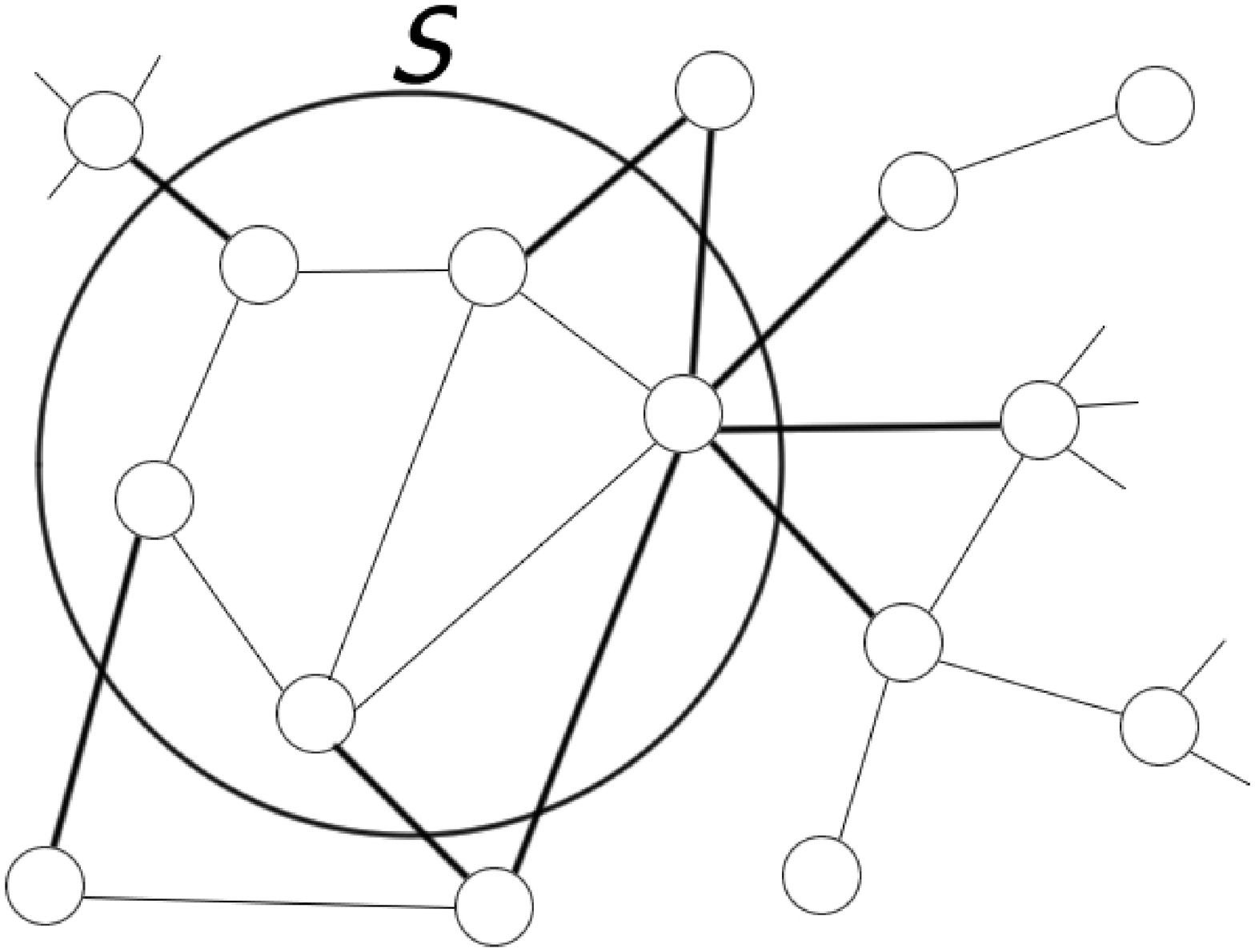}
\end{center}
\caption{The isoperimetric problem consists in finding the subset
$S$ such as to minimize the ratio $h_{G}(S)$ in Eq. (11). In this
schematic picture, a given subset S is selected as the set of all
network vertices contained inside the circle in the figure. Its
cardinality is given by the number of nodes in the circle, $|S|=5$
and its boundary is given by the number of edges (represented in
bold) which connect the subset $S$ to the rest of the network,
$\mathcal{D}(S)=9$.}
\end{figure}

As $\lambda_N$ is found to be almost independent from the
correlation coefficient $r$ (see Fig.1(c)), we will focus now on
estimating the effects of correlation on the eigenvalue $\lambda_2$,
the parameter known as \emph{algebraic connectivity of graphs}
\cite{Mohar1}. Specifically, to estimate an upper bound for
$\lambda_2$, we use the following {\em Cheeger inequality} from
graph theory \cite{Cheeg}:
\begin{equation}
\lambda_2\leq h_G, \label{eq:CI}
\end{equation}
with $h_G$, the Cheeger constant of a graph, defined as follows
\cite{Cheeg}:
\begin{equation}
h_G=\min_S h_G(S). \label{minimization}
\end{equation}
For a given subset of vertices, say $S$, $h_G(S)$ is the quantity
given by:
\begin{equation}
h_G(S)=\frac{{\mathcal D}(S)N}{|S|(N-|S|)},  \label{eq:hG}
\end{equation}
where ${\mathcal D}(S)$ is the number of edges in the boundary of
$S$ and $|S| < \frac{N}{2}$, is the number of vertices in $S$.

The problem of finding $h_G$, known as the isoperimetric problem in
graph theory, is illustrated by means of a representative example in
Fig. 3. Note that, given a graph of order $N$, there are
$\sum_{i=1}^{N/2} C(N,i)$ possible ways of choosing the subset $S$,
where $C(a,b)$ is the number of combinations of $a$ elements in $b$
places, and thus the problem of choosing $S$ such as to minimize
$h_G(S)$ is NP-hard.

We will show that (\ref{eq:CI}) can be successfully used to compute
an upper bound on $\lambda_2$. To overcome the limitations due to
the computation of the subset $S$ that minimizes $h_G(S)$, we will
follow a stochastic approach in order to estimate $h_G(S)$, starting
from the available information we have on the network. We will
assume that the noticeable features of the network are only the
degree distribution and the correlation specified; all other aspects
being completely random.

Note that, by considering as equivalent all the vertices
characterized by the same degree, the original problem in Fig. 3,
can be converted in the other (simpler) one shown in Fig. 4. In the
picture, the subset $S$ includes two vertices of degree $3$, two
vertices of degree $4$ and one vertex of degree $7$. In terms of the
nodes degrees the original problem, consisting of finding the
optimal set of vertices, can then be re-interpreted as that of
finding the optimal mix of degrees of nodes in $S$.

\begin{figure}[t]
\begin{center}
\includegraphics[width=7cm]{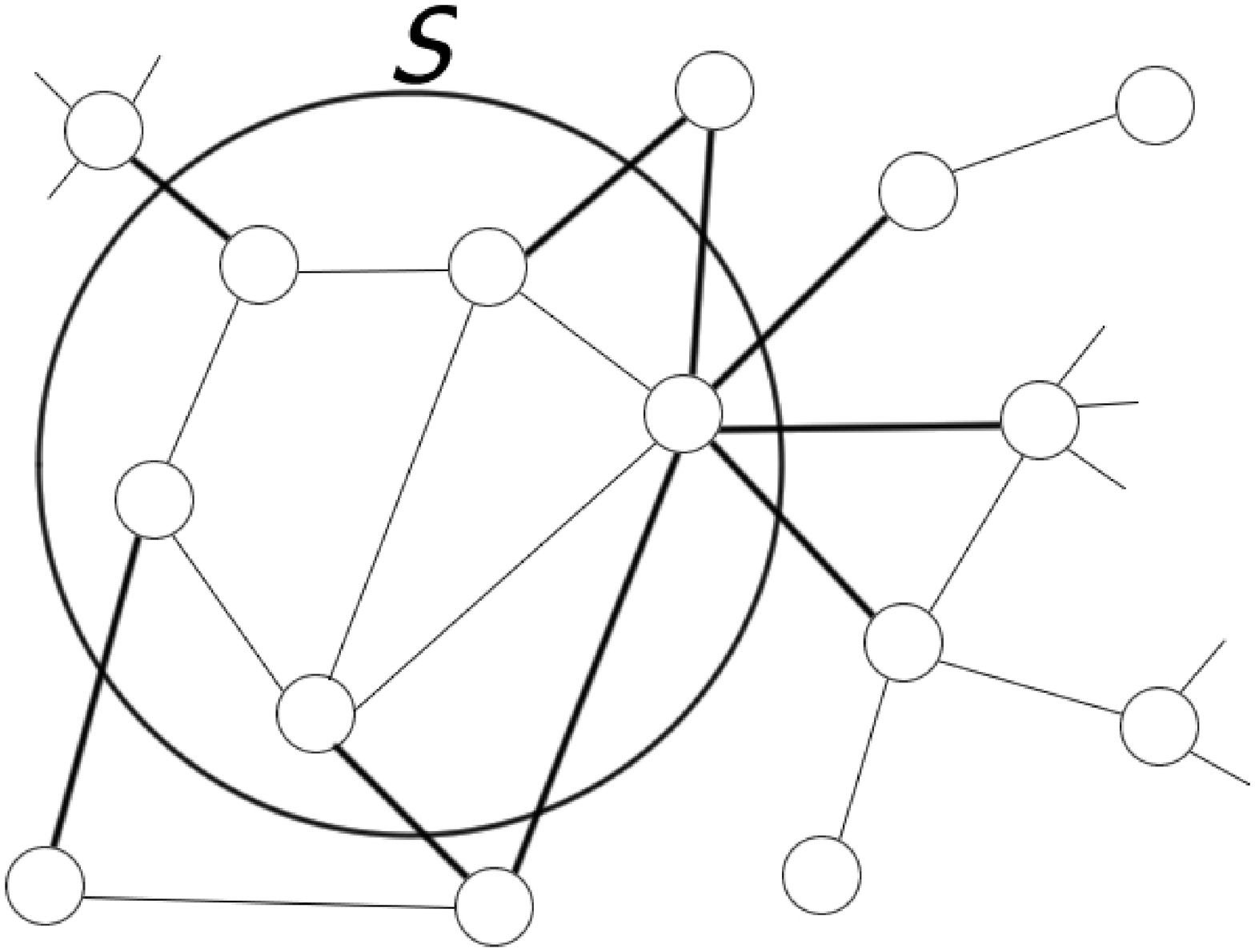}
\begin{picture}(0,0)(0,0)
\put(-132,106){ $\bf{4}$} \put(-185,68){ $\bf 3$}
\put(-164,106){\small $\bf 3$} \put(-159,34){ $\bf 4$}
\put(-97,82){\small ${\bf 7}$}
\end{picture}
\end{center}
\caption{The problem described in Fig.2 has been converted in the
one represented in this picture. 
Here, the subset $S$ is identified by the degrees of the vertices
inside it, which are two nodes of degree $3$, two nodes of degree
$4$ and one node of degree $7$. Note that, in the original
formulation, the problem was of choosing \emph{the vertices} to be
included in $S$ in order to minimize $h_G(S)$. Here we have restated
it by considering only \emph{the degrees} of the vertices to be
included in $S$, in terms of the vector $\bf{y}$,
such as to minimize $h_G({\bf y})$. 
}
\end{figure}

Then, we can give a full characterization of a randomly chosen
subset $S$ in terms of the number of nodes in it, say $x_i$,
belonging to each class ${\mathcal{C}}_i$ ($i=1,2...,p$). Let us
term as $y_i=\frac{x_i}{n_i}$ the fraction of nodes in $S$ drawn
from each class ${\mathcal{C}}_i$ ($i=1,2...,p$). Then problem
(\ref{minimization}) becomes that of finding the combination
${\mathbf{y}}={({y_1, y_2, ..., y_p})^{T}}$ such as to minimize
$h_G({\bf y})$, i.e. $\min_{\bf y} h_G({\bf y}, r)$.
 Then, since under this formulation nodes with the same degree are all clustered in the same class, the complexity of the
problem is reduced to approximately $\sum_{i=1}^{N/2} C(N,i)/
\Pi_{j=1}^{k_{max}} n_j$. It is also worth noting here that the
subset $S$ is not supposed to satisfy any particular condition, not
even of being connected.

Now, we observe that the number of edges in the boundary, say
${\mathcal D}(S)$, is given by the total number of edges starting
from the vertices in $S$, less the ones, say ${\mathcal
I}(S)$, that are contained in $S$, i.e. having both endpoints in $S$. 
Thus we can estimate $\mathcal{I}$ and ${\mathcal D}$ as follows:
\begin{eqnarray*}
& {\mathcal I}(S) = {\mathcal I}({\mathbf{y}}, r) =
{\mathcal{E}} {\mathbf{y}}^T E \mathbf{y}, \\
& {\mathcal D}(S)={\mathcal D}({\mathbf{y}}, r)={\mathbf{x}}^T
{\mathbf{k}} -2 {\mathcal I}(S)= {2 \mathcal{E} (\mathbf{y}}^T
{\mathbf{q}} - {\mathbf{y}}^T E \mathbf{y}),
\end{eqnarray*}
where ${\mathbf{x}}={({x_1,  x_2, ..., x_p})^{T}}$ and $\mathcal E$
is the total number of edges in the network.
%

Therefore $h_G(S)$ becomes:
\begin{equation}
\label{hGS} h_G({\mathbf{y}}, r)=\frac{{2 \mathcal{E} (\mathbf{y}}^T
{\mathbf{q}} - {\mathbf{y}}^T E {\mathbf{y}})N}{({\mathbf{n}}^T
{\mathbf{y}}) (N-{\mathbf{n}}^T {\mathbf{y}})},
\end{equation}
under the constraint that ${\mathbf{n}}^T {\mathbf{y}} <N/2$, where
${\mathbf{n}}$ is the vector $({n_1,  n_2, ..., n_p})^{T}$. A
numerical optimization algorithm can then be used to find the subset
$S$ that minimizes $h_G(\mathbf{y},r)$ in terms of $y_1, y_2, ...,
y_p$ (and subsequently $r$) and, in turns, an upper bound for
$\lambda_2$.

Also, from (\ref{EE}) and (\ref{hGS}), we get:
\begin{equation}
\label{proc}
\frac{\partial{h_G(S)}}{\partial{r}}\propto\frac{\partial{{\mathcal
D}(S)}}{\partial{r}}= 
-2 {\mathcal E}\sigma^2_q {({\mathbf{y}}^T \mathbf{m})}^2 \leq 0.
\end{equation}
Since, for any vector $\mathbf y$, (\ref{proc}) is satisfied, then
we have that $\frac{\partial{h_G}}{\partial{r}} \leq 0$. Therefore,
we can predict analytically that $h_G$ and hence $\lambda_2$ will be
decreasing as the degree correlation is increased and, as a
consequence, the eigenratio $\lambda_N \over \lambda_2$ will
increase for higher values of the correlation coefficient.


Another interesting inequality in spectral geometry is due to Mohar
\cite{Mohar}:
\begin{equation}
\label{eqM} \lambda_2 \geq k_{max}-\sqrt{{k_{max}}^2-{h'_{G}}^2},
\end{equation}
where $h'_G=\min_{S}{\frac{{\mathcal D}(S)}{|S|}}={2 \mathcal{E}
(\mathbf{y}}^T {\mathbf{q}} - {\mathbf{y}}^T E
{\mathbf{y}})/({\mathbf{n}}^{T} \mathbf{y})$. Using (\ref{eqM}), we
can also get a lower bound on $\lambda_2$. Then following an
approach similar to the one used to compute the upper bound, it is
easy to show that the lower bound in (\ref{eqM}) has to decrease
with $r$ (note that when making the correlation change, the degree
distribution is fixed and thus, $k_{max}$ cannot vary with $r$).
Since both the upper and the lower bounds have to decrease with $r$,
$\lambda_2$ is also expected to have the same trend.
%
%
%

\subsection{Disassortativity as an emerging property}

The main result of the derivation presented above is the finding
that disassortative networks synchronize better. We wish now to
assess whether negative degree correlation can be thought of as an
emerging property of networks with an assigned degree distribution
in order to improve their synchronization.
 In particular, we noticed that the variation
of the degree correlation affects mainly the Laplacian eigenvalue
$\lambda_2$. Thus, we shall seek to find if varying the correlation,
while keeping the degree distribution fixed, is indeed a good way of
optimizing the network synchronizability properties.

\begin{figure}[h]
\begin{center}
\includegraphics[width=13cm]{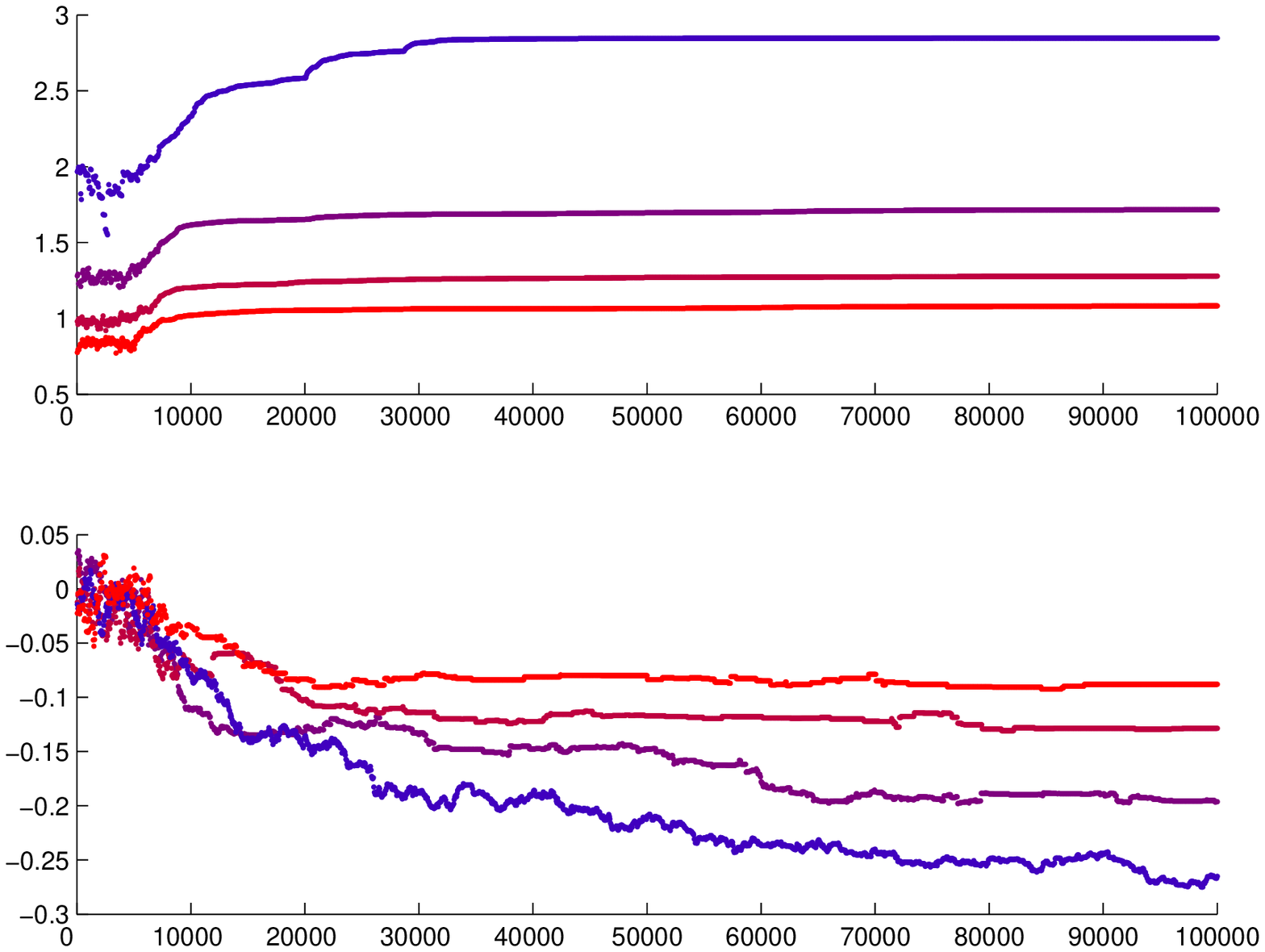}
\begin{picture}(0,0)(0,0)
\put(-345,63){\small ${r}$} \put(-345,210){\small $\lambda_2$}
\put(-168,-10){\small ${t}$} \put(-168,129){\small $t$}
\put(-100,15){\small $\gamma=2$} \put(-40,30){\small $\gamma=3$}
\put(-105,55){\small $\gamma=4$} \put(-75,80){\small $\gamma=5$}
\put(-50,160){\small $\gamma=5$} \put(-110,185){\small $\gamma=4$}
\put(-65,205){\small $\gamma=3$} \put(-75,248){\small $\gamma=2$}
\end{picture}
\end{center}
\caption{\label{fig:annealing} \small Effects of the maximization of
$\lambda_2$ on the correlation parameter $r$ defined in [11] for
different scale-free networks with $\gamma=[2,3,4,5]$.}
\end{figure}
This might be a solution to a classical problem in graph theory
optimization which is the construction of expander graphs, i.e.
highly efficient communication networks, characterized by high
values of $\lambda_2$ \cite{Al:Exp} (and therefore low values of the
eigenratio $R$).

We will use a simulated annealing meta-heuristic technique, to solve
the problem of maximizing $\lambda_2$ while keeping unchanged the
network degree distribution. To this aim, given a network with a
certain degree distribution, we will perform the following iterative
procedure. At each step, the endpoints of a randomly selected pair
of edges are exchanged if
$\exp{(\frac{-\Delta(\lambda_2)}{T})}>z$, where $z$ is an 
uniformly distributed random variable between 0 and 1,
$\Delta(\lambda_2)$ is the variation achieved in the objective
function $\lambda_2$ before and after the execution of the move and
$T$ is a control parameter, which similarly to the original
formulation of the simulated annealing procedure is known as the
system 'temperature'. As the algorithm runs, the temperature $T$ is
decreased according to an \emph{exponential cooling scheme} (see
\cite{Ki:Ge:Ve} for further details).

 As shown in Fig. \ref{fig:annealing}, while trying to maximize
$\lambda_2$, we observe the spontaneous emergence in the network of
interest of negative degree correlation, namely an increase of its
disassortativity.

Thus, following an entirely different approach, we come to the
conclusion that if the degree distribution of a given network is
fixed, then to improve its synchronizability one has to introduce
negative degree correlation among its node. This suggests that in
evolutionary biological networks of nonlinear oscillators,
disassortativity might be an emerging property necessary to
optimize the synchronization process.

\section{Weighed Networks}

\begin{figure}[t]
\begin{center}
\includegraphics[width=13cm,height=8cm]{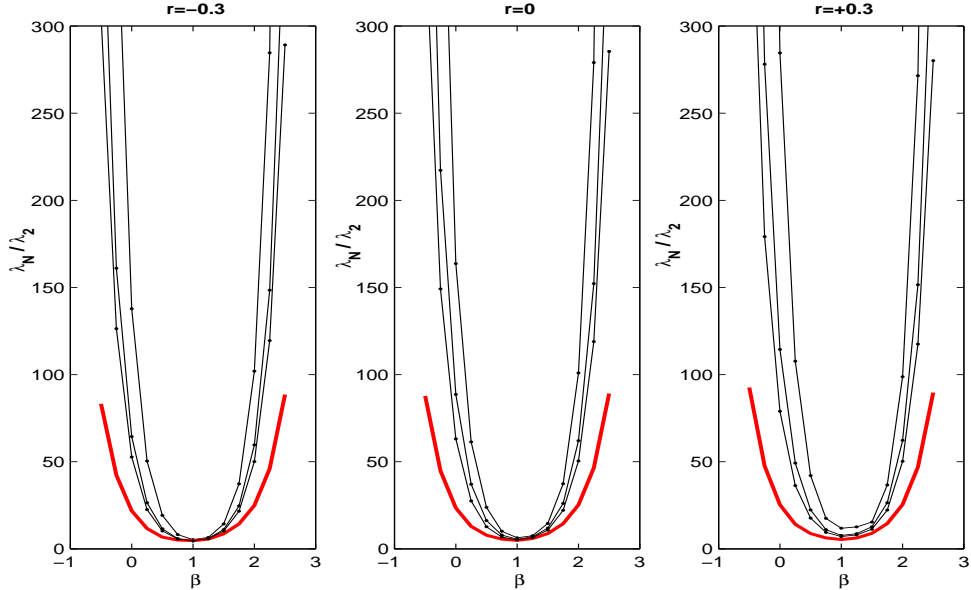}
\end{center}
\caption{\label{fig:lambda_beta} \small Plots of
$\lambda_N/\lambda_2$  as varying $\beta$ for scale free networks
with different values of the power-law exponent
$\gamma=[2.5,3,3.5]$. On the left plot we show negatively correlated
networks ($r=-0.3$), on the right positively correlated ones
($r=+0.3$), in the center, networks characterized by absence of
degree correlation ($r=0$).  The thick lines represent the behavior
of the eigenratio for random networks ($\gamma=\infty$). $N=10^3$
nodes, $\mathcal{E}=4 \cdot 10^3$.}
\end{figure}

In this section we shall extend the results presented above to the
case of weighed topologies. Interestingly in \cite{paradox,Bocc1},
it has been shown that an appropriate choice of weights over the
network links, may improve considerably the network
synchronizability. In \cite{Bocc1}, it has been shown that such
improvement is highest if the weights are set to be proportional to
a particular topological property at the network links, termed as
\emph{load} or \emph{betweenness centrality}.
\begin{figure}[h]
\begin{center}
\includegraphics[width=13cm]{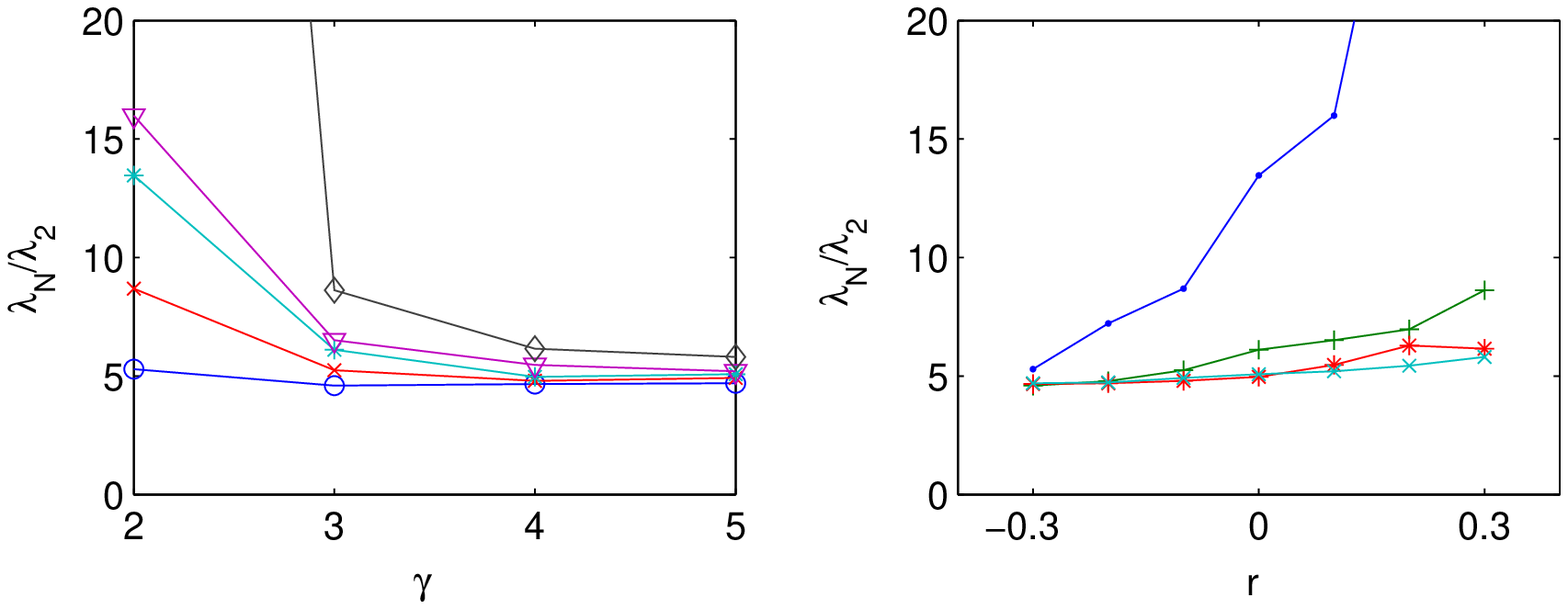}
\end{center}
\caption{\label{fig:R_gamma} \small Synchronizability of degree
correlated networks in the optimal regime ($\beta=1$). Left figure:
behavior of the eigenratio $\lambda_{N}/\lambda_{2}$, as a function
of $\gamma$. 
$N=10^3$, $\mathcal{E}=4 \cdot 10^3$. Legend  is as follows: $r=-0.3
(o)$, $r=-0.1 (\times)$,  $r=0 (*)$, $r=0.1 (\triangledown)$, $r=0.3
(\diamondsuit)$. Right side: behavior of the eigenratio
$\lambda_{N}/\lambda_{2}$ as a function of the correlation
coefficient $r$, for $\gamma$ varying from 2 to 5. $N=10^3$,
$\mathcal{E}=4 \cdot 10^3$. Legend is as follows: $\gamma = 2
(\cdot)$ , $\gamma = 3 (+)$, $\gamma = 4 (*)$, $\gamma = 5
(\times)$.}
\end{figure}
\begin{figure}[h]
\begin{center}
\includegraphics[width=13cm]{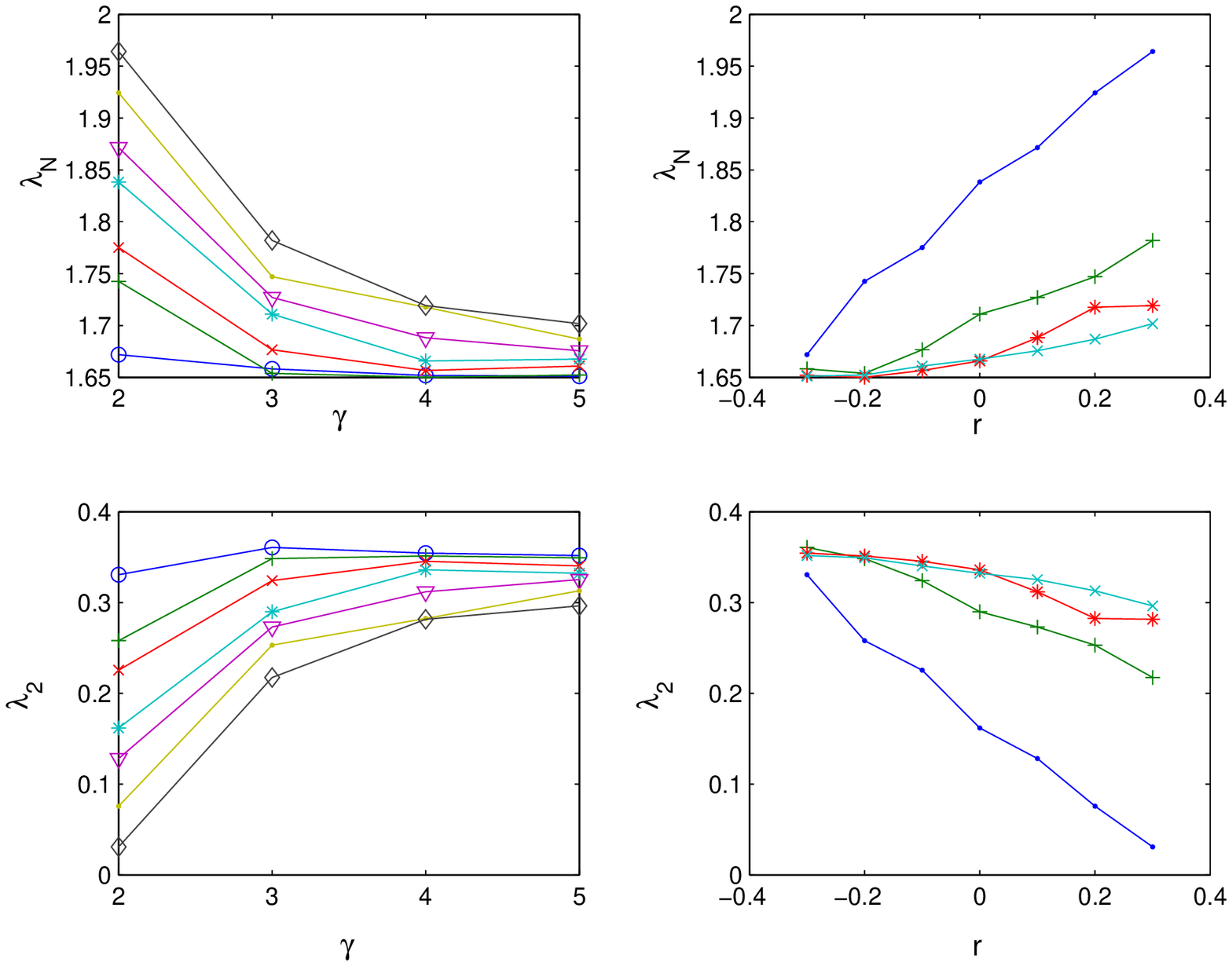}
\begin{picture}(0,0)(0,0)
\put(-250,250){\small ${(a)}$} \put(-250,60){\small ${(c)}$}
\put(-120,250){\small ${(b)}$} \put(-120,60){\small ${(d)}$}
\end{picture}
\end{center}
\caption{\label{fig:Eigminmax} \small Synchronizability of degree
correlated networks in the optimal regime ($\beta=1$). $N=10^3$
nodes, $\mathcal{E}=4 \cdot 10^3$. Behavior of the highest
eigenvalue $\lambda_N$ (a) and of the second lowest one
$\lambda_2$ (c), as functions of the correlation coefficient $r$,
for $\gamma=[2,3,4,5]$. Behavior of the highest eigenvalue
$\lambda_N$ (b) and of the second lowest one $\lambda_2$ (d), as
functions of $\gamma$, for $r=[-0.3,0,+0.3]$. Legend in (a) and
(c) is as follows: left side, $r=-0.3 (\circ)$, $r=-0.2 (+)$,
$r=-0.1 (\times)$, $r=0 (*)$, $r=0.1 (\triangledown)$, $r=0.2
(\cdot)$, $r=0.3 (\diamondsuit)$. Legend in (b) and (d) is as
follows: $\gamma = 2 (\cdot)$ , $\gamma = 3 (+)$, $\gamma = 4
(*)$, $\gamma = 5 (\times)$.}
\end{figure}

We have seen sofar that scale-free networks can synchronize better
as the degree distribution exponent $\gamma$ is increased and the
degree correlation coefficient $r$ is decreased. In
\cite{Mo:Zh:Ku04}, it was shown that synchronizability of such
networks can be further enhanced by an appropriate choice of the
network coupling weights, that takes into account exclusively
topological information. Specifically, it was found that in order
to enhance the synchronizability of heterogeneous networks, the
coupling matrix should have the form:
\begin{equation}
\label{coupling} \mathcal{L}=D^{-\beta}(D-A),
\end{equation}
where the strength of the coupling, for each edge, is scaled by a
power (with exponent $\beta$) of the degree of the starting node.
Note that, by tuning the parameter $\beta$, it is possible to vary
the strength of the coupling from high to low degree vertices and
viceversa. Also,  from (\ref{coupling}), $\mathcal{L}$ can be
rewritten as $D^{-\beta/2}(D-A)D^{-\beta/2}$  and thus, even if
$\mathcal{L}$ is not symmetric, its spectrum is always real,
whatever the value of $\beta$.

In \cite{Mo:Zh:Ku04}, in the case of heterogeneous networks, an
optimal coupling was found to be  $\beta=1$ in (\ref{coupling}).
It is worth noting that this leads to asymmetric coupling. Take
for example the link between an high degree and a low degree
vertex; a coupling of the form (\ref{coupling}), with $\beta=1$,
results in the low degree node having a much higher influence on
the higher degree one than
viceversa. 

Generally speaking, the case of $\beta=1$, with the weights of the
incoming links at each vertex summing to -1 (so that
$\mathcal{L}_{ii}=1 \quad \forall i$), is of particular interest in
that all the networks satisfying this constraint, become directly
comparable among themselves, in terms of their spectral properties \cite{Bocc1} (see also \cite{Sorr:db06}.) 
In fact, according to the Gerschgorin's circle theorem, all the
eigenvalues ($\lambda_j= \lambda_j^r+ j \lambda_j^i, \quad
j=1,2,...,N$) of these network are known
 to belong to a bounded
region of the complex plane, and specifically the circle of radius
1, centered at 1 on the real axis \cite{Bocc1}. Thus all the
$\lambda_j^r$, are constrained to belong to the interval [0,2] (i.e.
$0=\lambda_1^r \leq \lambda_2^r \leq ... \leq \lambda_{N}^r \leq 2$)
and the $\lambda_j^i$ to the interval [-1,1], $j=1,2,...,N$ (in the
case of interest here of eq. (\ref{coupling}), $\lambda_j^i=0  \quad
\forall j$). Moreover the case of $\beta=1$ has been shown to
represent an optimum in terms of the network synchronizability,
independently from the form of its degree distribution.

Moreover in \cite{Mo:Zh:Ku04}, it was claimed that in the optimally
coupled network ($\beta=1$), synchronizability is solely determined
by the average degree, while heterogeneity  in the network
connectivity does not affect synchronization.  However, as shown in
Sec.~4.1, we observe that in the case of assortative mixing, this is
not true anymore; specifically in the particular case of $r=0.3$
shown in the right panel of Fig.~\ref{fig:lambda_beta}, homogenous
networks are found to behave better than heterogenous ones, even in
the optimal case of $\beta=1$.

\subsection{Effects of degree correlation}
Here, we will show that, assuming a coupling of the form
(\ref{coupling}), the presence of degree correlation does not alter
the optimal value for $\beta$ and better synchronizability
properties are still detected when $\beta=1$ as was observed   in
the uncorrelated case. Moreover, the presence of degree correlation,
unlike heterogeneity, does continue to have an effect on the network
synchronizability, even in the optimal regime. (Namely, as in the
case of unweighed coupling, disassortative mixing always enhances
network synchronization).


In Fig. \ref{fig:lambda_beta} we have reported the behavior of the
eigenratio as varying $\beta$ in networks with different degree
correlation properties (characterized by $r=[-0.3,0,+0.3]$). Our
numerics confirm that the eigenratio $R=\lambda_N/\lambda_2$ is
characterized by a peaked minimum at $\beta \simeq 1$, independently
from the value of $r$ (i.e. also when $r \neq 0$). Note that in the
case of uncorrelated networks, when $\beta=1$, the eigenratio is
quite insensible to the specific form of the degree distribution, as
claimed i.e. in
\cite{Mo:Zh:Ku04}. Nonetheless, as $r$ increases, 
the eigenratio appears to be more sensible to variations in the
degree distributions, even in the optimal case of $\beta=1$.
Specifically, as shown in Fig.~\ref{fig:R_gamma}, in such a case the
minimum is more pronounced for higher values of $\gamma$ (i.e. the
paradox of heterogeneity is still present even at $\beta=1$).

Differently from the case of unweighed networks, we notice that in
Fig.~\ref{fig:Eigminmax} $~\lambda_N$ is now sensible to variations
in $r$, when compared to $\lambda_2$; thus the behavior of the
eigenratio $R$ as $r$ varies, cannot be explained only in terms of
the second smallest eigenvalue but is a combination of variations of
both the eigenvalues (see Fig. \ref{fig:Eigminmax}(a) and
\ref{fig:Eigminmax}(c)). The behavior of $\lambda_2$ and $\lambda_N$
under variations of $\gamma$ for different values of $r$ is shown in
Fig.~\ref{fig:Eigminmax}(b)-(d). Namely, when increasing $\gamma$,
we observe a rise in $\lambda_2$ and at the same time, a decrease in
$\lambda_N$, leading to an overall decrease of the eigenratio $R$
and therefore better synchronizability.  Similarly, when $r$ is
varied from the disassortative ($r<0$) to the assortative ($r>0$)
regime, we observe both a decrease in $\lambda_2$ and an increase in
$\lambda_N$, leading to a rise in the eigenratio $R$.

Observe that in the case of the normalized Laplacian
(\ref{coupling}) considered here, the following relationship from
graph theory is valid \cite{ChungBOOK}:

\begin{equation}
\frac{{h''}_G^2}{2} <  \lambda_2 \leq 2 {h''}_G,
\end{equation}
where $h''_G= \min_S h''_G(S)$ and $h''_G(S)$ is defined as:

\begin{equation}
h''_G(S)=\frac{{\mathcal{D}(S)}}{\min{(vol(S), vol(\bar{S})})},
\end{equation}
with $vol(S)=\sum_{i \in S} k_i$. Note that in this case $h''_G(S)$
can be estimated as:

\begin{equation}
\label{hG2} {h''}_G({\mathbf{y}}, r)=\frac{{ (\mathbf{y}}^T
{\mathbf{q}} - {\mathbf{y}}^T E {\mathbf{y}})}{\min({\mathbf{y}}^T
{\mathbf{q}},\quad (\bf{1}-{\mathbf{y}}^T) {\mathbf{q}})}.
\end{equation}

Then it is sufficient to observe that, for each $S$, the denominator
in
(\ref{hG2}) is a quantity independent from $r$, 
and it becomes straightforward to obtain the existence of an upper
and a lower bound for $\lambda_2$, decreasing with $r$.
Unfortunately notwithstanding this,  estimating the effects of $r$
on $\lambda_N$ resulted to be a non-trivial task and therefore it
remains an open problem. An important feature seems to be the
symmetry between the behavior of $\lambda_2$ and $\lambda_N$
reported in Fig. ~\ref{fig:Eigminmax}(a)-(c) and
~\ref{fig:Eigminmax}(b)-(d).

From the numerics shown in Fig.~\ref{fig:Eigminmax}, we observe a
clear advantage of introducing negative degree correlation within
the network (in terms of both $\lambda_N$ and $\lambda_2$), when
$\beta=1$. Thus we expect to observe an enlargement of the
synchronization interval $I_{\sigma}$, for both higher and lower
values of $\sigma$ as decreasing $r$. Moreover, since this happens
for whatever a choice of $\beta$, we find that disassortative mixing
always enhances synchronization.

The effects of correlation on the eigenratio $R$ in the case of
 $\beta=1$ (i.e. optimal synchronizability) is fully
illustrated in Figs.~\ref{fig:R_gamma} and \ref{fig:Eigminmax}.
Specifically from Figs.~\ref{fig:R_gamma} and \ref{fig:Eigminmax},
we observe that negatively correlated networks continue to show
higher synchronizability, for every value of the exponent $\gamma$.
This indicates that the effects of correlation on $R$ are not
dampened by the presence of an appropriate coupling of the form
(\ref{coupling}). Fig. \ref{fig:R_gamma} also shows a non-negligible
dependence of the eigenratio on the exponent of the power-law
distribution $\gamma$, indicating that heterogeneity continues to
affect the synchronizability of weighed correlated networks,
especially in the cases where $r>0$ (as also shown in the right
panel of Fig. \ref{fig:lambda_beta}).

\section{Effects of degree correlation on the synchronization dynamics}
Now, we shall seek to investigate the effects of variable degree
correlation on the dynamics of networks of oscillators. Specifically
in what follows we will evaluate the behavior of dynamical networks
of (i) identical and (ii) non identical oscillators.

As an example of identical oscillators, we consider a network of
coupled chaotic R\"ossler oscillators, connected as in (1). In
\cite{IJBC_NDES}, we have already reported numerical results showing
the synchronization dynamics of such networks in the case where
$h(x)=Ix$, with $I$ being the 3-dimensional identity matrix. Therein
we observed that negative degree correlation does indeed have
beneficial effects on the network synchronization. Specifically, our
preliminary results confirmed that negative degree correlation is
able to enhance the network synchronization dynamics.

Here we suppose the dynamics at each node $i$ to be described by the
following vector field, ${\bf{v_i}} =(x_i,y_i,z_i)$:
\begin{eqnarray*}
   &\dot x_i= & - y_i\; -\; z_i-  \sigma  \sum_{j=1}^{N} {\mathcal L_{ij}} x_j\\
   &\dot y_i= & x_i\;+\;0.165 y_i\\
   &\dot z_i= & 0.2+(x_i-10)z_i- \sigma \sum_{j=1}^{N} {\mathcal L_{ij}} z_j\\
\end{eqnarray*}
where  $\sigma$ is the tunable strength of the coupling. Note that,
according to this scheme, we have chosen the R\"ossler to be coupled
through the variables $x$ and $z$. The reason is that under this
assumptions, the MSF results to be negative in a bounded range of
values of the parameter $\alpha$, yielding that the coupling
$\sigma$ should be neither too low, nor too high.
\begin{figure}[h]
\centering
\includegraphics[width=6cm]{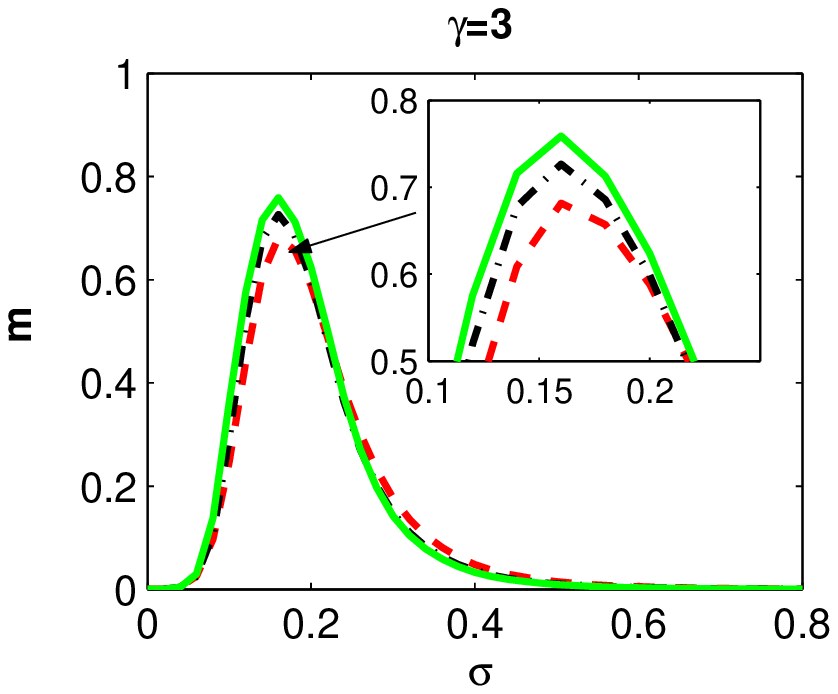}
\includegraphics[width=6cm]{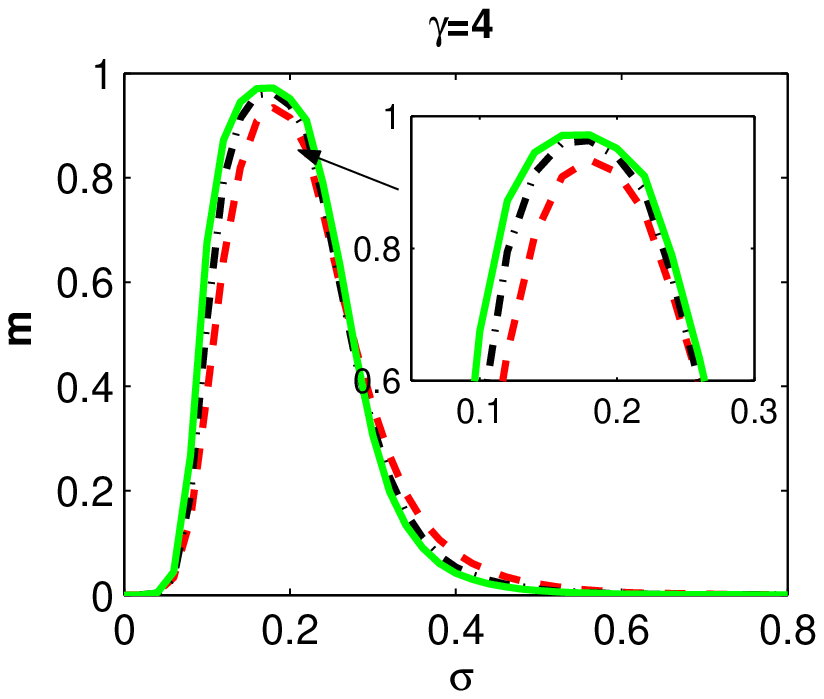}
\includegraphics[width=6cm]{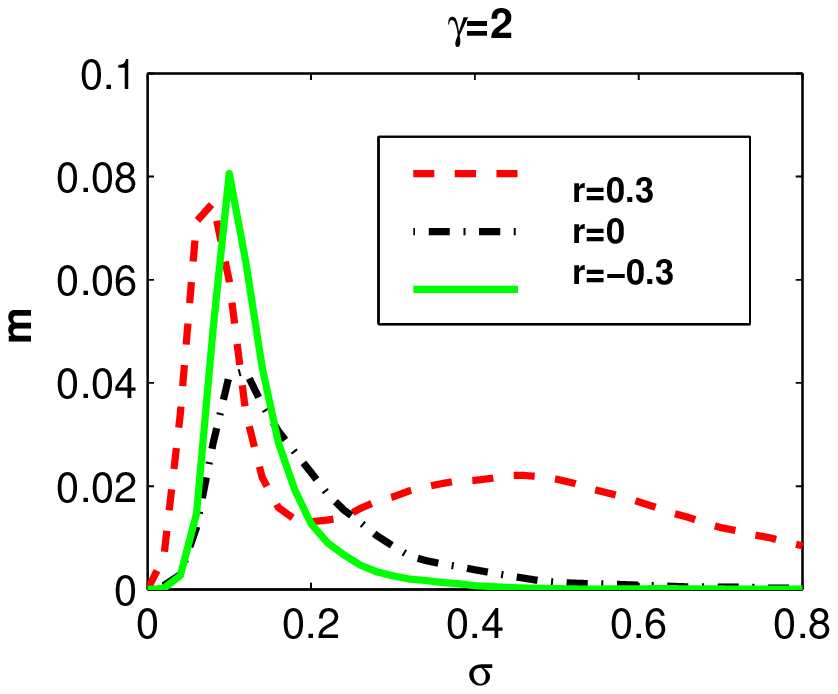}
 \caption{Synchronization dynamics over unweighed networks ($\beta=0$). The
plots show the order parameter at regime $m$ for networks
characterized by different $\gamma$ (in each subplot) and $r$
(differently colored lines) for increasing values of the coupling
gain $\sigma$; $N=10^3$, $\mathcal{E}=4 \cdot 10^3$. Note the
different scale on the y-axis in the lower figure.\label{campanaNN}}
\end{figure}
\begin{figure}[h]
\centering
\includegraphics[width=13cm]{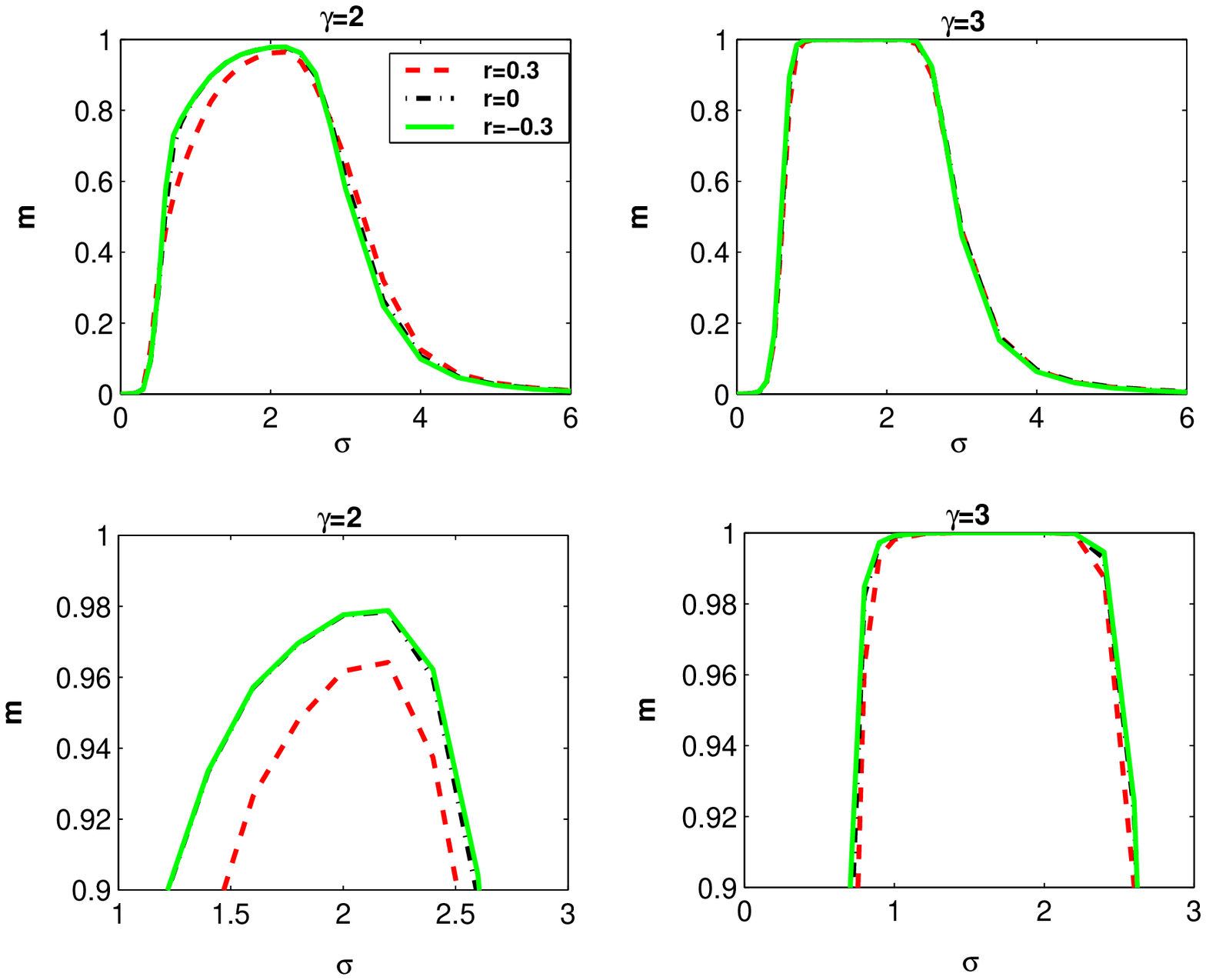}
 \caption{Synchronization dynamics over weighed networks ($\beta=1$).
 The plots show the
order parameter at regime $m$ for networks characterized by
different $\gamma$ (in each subplot) and $r$ (differently colored
lines) for increasing values of the coupling gain $\sigma$;
$N=10^3$, $\mathcal{E}=4 \cdot 10^3$. The lower plots are
enlargements of the upper ones. In the right-low panel, it is
interesting to observe how the range of values of $\sigma$ for
which synchronization is achieved (at $m=1$) increases as $r$ is
reduced from 0.3 to -0.3. \label{campanaN}}
\end{figure}

We have performed simulations over scale-free complex networks,
with variable power-law exponents, characterized by several values
of the degree correlation coefficient (typically in the range $r
\in [-0.3,+0.3]$).

In order to have a measure of the overall network synchronization,
we introduce
the order parameter $m(t) \in [0,1]$, defined as follows (see also \cite{Hild1}): 
\begin{equation}
  \label{m}
  m(t)={1\over N(N-1)}\sum_{j\ne i,i,j=1}^N \Theta(\delta - d_{ij}(t)),
\end{equation} 
where $\Theta(x)$ is the Heavyside function, i.e. $\Theta(x)=1$ if
$x\ge 0$ and $\Theta(x)=0$ otherwise. The parameter $\delta$ is a
small number to account for the finite numerical accuracy, (here we
have chosen $\delta = 10^{-2}$), so that two states in phase space
lying inside a sphere of radius $\delta$ are considered as mutually
being synchronized. The parameter $m(t)$ gives the fraction of pairs
of elements $(i,j)$ which are synchronized at time $t$ (i.e.,
$d_{ij}\leq \delta$). This fraction is equal to unity if all
possible pairs are synchronized and zero if no pair is synchronized,
with intermediate
values $0<r<1$ indicating 
partial synchronization. In what follows we will look at the
asymptotic average value of $m(t)$, say $m$, as function of the
coupling strength $\sigma$.

 Specifically, in Figs. \ref{campanaNN}
and \ref{campanaN}, a double phase transition is observed: the
first one from a non-synchronous to a synchronous regime as the
coupling strength is increased; the second one from the
synchronous to a non-synchronous phase as the coupling strength is
further increased and exceeds a critical value.

At first, we suppose the network topology to be unweighed, i.e. the
coupling is identical over all the edges and equal to 1. In order to
explore the effects of the network structure (degree distribution
and degree correlation) on the dynamical synchronization process, we
have performed several numerical simulations, the main results being
illustrated in Fig. \ref{campanaNN}. The asymptotic value of the
order parameter $m$, has been computed for networks with different
topological features. We have compared scale free networks
characterized by different power-law exponents and we found that, as
can be observed by comparing the plots in Fig. \ref{campanaNN},
homogenous networks (characterized by higher values of $\gamma$)
synchronize better than heterogenous ones (note the different scale
on the y-axis in the lower plot, where heterogeneous scale-free
networks, with $\gamma=2$ cannot be synchronized in practice for any
value of $\sigma$). This indicates that an high heterogeneity in the
degree distribution can result in being very harmful to the overall
synchronization \cite{Ni:Mo}. Moreover negative degree correlation
is seen to enhance the network synchronizability, for all the values
of $\gamma$ considered (excluding those for which synchronization
cannot be achieved, e.g. $\gamma=2$ in Fig. 9, where $m \ll 1$).
This is consistent with the predictions based on the eigenvalue
analysis provided in Sec. 3 and confirms the beneficial role played
by negative degree correlation with respect to networks
synchronization even in terms of synchronization dynamics.

The case of weighed topologies, (i.e. when $\beta=1$) is shown in
Fig. \ref{campanaN}. Again, negative degree correlation is observed
to enhance the network synchronization, at high values of the order
parameter when $m \approx 1$. Moreover, as shown in Fig.
\ref{campanaN}, the network synchronization is reduced as the
heterogeneity in the degree distribution is increased (i.e for lower
values of the exponent $\gamma$). This indicates that, even in the
optimal regime where $\beta=1$, networks characterized by variable
degree distributions, behave differently in terms of their
synchronization dynamics (practically we observe an analogous
phenomenon as in the case of the unweighed topologies). Also, it is
interesting to observe in Fig. \ref{campanaN} that the range
$I_{\sigma}$ of values of $\sigma$ for which synchronization is
completely achieved (at $m=1$), increases by reducing $r$ from 0.3
to -0.3.

It is worth noting here that, by comparing Figs. \ref{campanaNN}
and \ref{campanaN}, weighed networks, in which the strength of the
coupling has been rescaled by the degree at each node, are more
synchronizable than unweighed ones (as also predicted by the
behavior of the eigenratio in Fig. \ref{fig:lambda_beta}). We wish
to emphasize that \emph{all the numerical results} provided in
this section, are in good agreement with the eigenvalue analysis
presented in Sec.3.

\section{Conclusions}
The structure of many real world networks is characterized by
non-trivial degree-degree correlation in the network connections.
This leads alternatively to disassortative mixing in biological and
technological networks and assortative mixing in social networks. In
this paper we have studied the effect of degree correlation on the
network synchronizability and synchronization dynamics.
Specifically, we studied the effects of correlation on the Laplacian
eigenratio, a parameter proposed in \cite{Ba:Pe02} as a measure of
the synchronizability of a network of coupled nonlinear oscillators.

We have found that disassortative mixing, which is typical of
biological and technological networks, plays a positive role in
enhancing network synchronizability. The numerical observations were
confirmed by the analytical estimates found for the Laplacian
eigenratio, which were shown to be well suited to describe the
observed phenomena.

Following \cite{Mo:Zh:Ku04}, we then analyzed the effects of
weighted and directed coupling of the form (\ref{coupling}) and
found that the presence of correlation continues to affect
synchronization with disassortative weighted networks synchronizing
better than assortative ones. We noticed that, even in the presence
of degree correlation, synchronizability seems to be optimal when
the strength of the coupling along each edge is made inversely
proportional to the degree of the starting node. Though a common
belief is that when such coupling is considered, heterogeneity
becomes unable to suppress synchronization, our numerics show that
an exception to this paradigm is represented by the case of
assortatively mixed networks, where the effect of the degree
distribution is found to be strongly enhanced.

Using a nonlinear optimization approach we found that negative
degree correlation is naturally attained by the network when the aim
is to minimize $\lambda_2$ and hence enhance synchronizability.
Thus, we conjectured that disassortative mixing has played the role
of a self organizing principle in leading the formation of many real
world networks as the Internet, the World Wide Web, proteins
interactions, neural and metabolic networks.

Finally, we investigated the synchronization dynamics of both
weighed and unweighed networks of identical R\"ossler oscillators,
confirming the theoretical results obtained in the paper. A more
general case of weighed degree correlated networks (with complex
spectrum of the Laplacian) will be discussed elsewhere
\cite{Sorr:db06}.

\end{document}